\documentclass[aps,prl,twocolumn,superscriptaddress,amsmath,amssymb]{revtex4-1}

\usepackage{graphicx}
\usepackage{multirow}
\usepackage{natbib}
\usepackage{array}
\usepackage{multirow,amssymb,amsbsy,amsmath,epstopdf}
\usepackage{color,soul}
\usepackage[colorlinks,citecolor=blue]{hyperref}

\definecolor{golden}{rgb}{0.8,0.6,0.1}

\newcommand{\xfrac}[2]{{#1}/{#2}}
\newcommand{\rfrac}[2]{{#1}/({#2})}
\newcommand{\beq}{\begin{equation}}
\newcommand{\eeq}{\end{equation}}
\newcommand{\erf}[1]{Eq.~(\ref{#1})}

\begin{document}

\title{Observing 
momentum disturbance in double-slit ``which-way'' measurements}
\author{Ya  Xiao}
\affiliation{CAS Key Laboratory of Quantum Information, University of Science and Technology of China, Hefei 230026, People's Republic of China}
\affiliation{Synergetic Innovation Center of Quantum Information and Quantum Physics, University of Science and Technology of China, Hefei 230026, People's Republic of China}

\author{Howard M. Wiseman}\email{h.wiseman@griffith.edu.au}
\affiliation{Centre for Quantum Dynamics, Griffith University, Brisbane, Queensland 4111, Australia}

\author{Jin-Shi Xu}\email{jsxu@ustc.edu.cn}
\affiliation{CAS Key Laboratory of Quantum Information, University of Science and Technology of China, Hefei 230026, People's Republic of China}
\affiliation{Synergetic Innovation Center of Quantum Information and Quantum Physics, University of Science and Technology of China, Hefei 230026, People's Republic of China}

\author{Yaron Kedem}
\affiliation{Department of Physics, Stockholm University, AlbaNova University Center, 106 91 Stockholm, Sweden}

\author{Chuan-Feng Li}\email{cfli@ustc.edu.cn}
\affiliation{CAS Key Laboratory of Quantum Information, University of Science and Technology of China, Hefei 230026, People's Republic of China}
\affiliation{Synergetic Innovation Center of Quantum Information and Quantum Physics, University of Science and Technology of China, Hefei 230026, People's Republic of China}

\author{Guang-Can Guo}
\affiliation{CAS Key Laboratory of Quantum Information, University of Science and Technology of China, Hefei 230026, People's Republic of China}
\affiliation{Synergetic Innovation Center of Quantum Information and Quantum Physics, University of Science and Technology of China, Hefei 230026, People's Republic of China}

\date{\today}

\begin{abstract}
Making a ``which-way" measurement (WWM) to identify which slit a particle goes through in a double-slit apparatus will reduce the visibility of interference fringes. There has been a long-standing controversy over whether this can be attributed to an uncontrollable momentum transfer. To date, no experiment has characterised the momentum change in a way that relates quantitatively to the loss of visibility. Here, by reconstructing the Bohmian trajectories of single photons, we experimentally obtain the distribution of momentum change, which is observed to be not a momentum kick that occurs at the point of the WWM, but nonclassically accumulates during the propagation of the photons. We further confirm a quantitative relation between the loss of visibility consequent on a WWM and the total (late-time) momentum disturbance. Our results emphasize the role of the Bohmian momentum in giving an intuitive picture of wave-particle duality and complementarity.

\end{abstract}

\maketitle


The single-particle Young's double-slit experiment is the quintessential example {of} the wave-particle duality of quantum mechanics~\cite{Heisenberg1927,Bohr1928}. If one performs a position measurement to determine which slit a quantum particle  traverses (particle-like property), the interference pattern (wave-like property) is damaged. The more ``which-way'' information one obtains, the lower the visibility of the interference fringes \cite{WZ1979,GY1988,JSV1995,Englert1996}. However, there has been a vigorous debate on whether the ``which-way" measurement (WWM) destroys interference by disturbing the momentum of the particle \cite{SEW1991,STCW1994,ESH1995,STCW1995,DNR1998,Wiseman1995}.

Opposite conclusions were obtained by two research groups. In 1991, Scully, Englert and Walther (SEW) proposed a WWM scheme to prove that one can perform a position measurement with sufficient precision to identify which slit the particle goes, without apparentely disturbing its momentum at all \cite{SEW1991}. They attributed the loss of visibility to the correlations between particles and the detectors. However, soon after, Storey, Tan, Collett and Walls (STCW) provided a general formalism  which appeared to show  that the detection of path information necessarily involves some momentum transfer to the particles \cite{STCW1994}.
A careful analysis~\cite{Wiseman1995} resolved this contradiction by showing that SEW and STCW were using different concepts of momentum transfer: `classical' and `quantum' respectively. That is, their analyses were in fact complementary. SEW's scheme could not be explained by a classical {\em  probability distribution} for momentum kicks, while the STCW theorem did correctly establish that there must be a nonzero {\em probability amplitude} for a momentum change of the expected size.

To study the paradigm of particle-wave duality in more depth, we need a more robust way to quantify the momentum disturbance. Neither SEW nor STCW gave such a measure for general situations. The difficulty is that we cannot unambiguously determine the momentum change to the quantum particle if the particle is not initially in a momentum eigenstate, which is the situation we face in a two-slit experiment. Bohmian mechanics, however, offers a way to solve this difficulty, as it posits that a particle has a definite position and momentum at all times, and hence follows a deterministic trajectory \cite{Bhom1,Bhom2}.
The Bohmian probability distribution for momentum transfer  was introduced in Ref. \cite{Wiseman1998} and showed to be well suited to characterising the momentum transfers in a WWM, both classical (immediate) and quantum (delayed) \cite{Wiseman1998}. It is a true probability distribution, and moreover can be experimentally observed using established techniques \cite{KBR2011,MRF2016,YX2017}. This means it is possible to experimentally explore the relation between the size of the momentum disturbance and the degree of visibility loss in a WWM.

In this work, we sent a triggered single photon through a birefringent double-slit apparatus and reconstructed its Bohmain trajectories using the technique of weak measurement \cite{Wiseman2007-2,Durr2009,KBR2011,MRF2016,YX2017}. Then we obtained the distribution of Bohmian momentum transfer to the particle in a WWM by comparing all the photon's trajectories in the free case and disturbed (WWM) case. We showed that the momentum change gradually accumulates during the propagation of the photons, which is negligible at short times. We further demonstrated the mean of the absolute value of the total (late-time) Bohmian momentum transfer $ \langle\vert p\vert\rangle^{B}_{T}$  to be larger than $  \rfrac{2\hbar}{\pi d} $, where $d$ is the center distance between the two slits. By implementing partial WWMs  experimentally, we found that this mean and the visibility satisfy the inequality $ \langle\vert p\vert\rangle^{B}_{T}\geq(1-V)\rfrac{2\hbar}{\pi d}$. That is, increasing the momentum disturbance to the particle in a WWM is observed to be accompanied by a decreasing of the visibility of the interference fringes, thus quantitatively demonstrating wave-particle duality.

According to Bohmian mechanics \cite{Bhom1,Bhom2}, an individual particle has a definite position and momentum; see also the Supplementary Information (SI). A Bohmian particle's momentum is determined by its position $x$. It can be obtained experimentally by performing a weak measurement of its momentum, post-selecting on finding it at position $x$,
and averaging the result over many repetitions. This yields

\begin{equation}\label{weak}
p(x) =  \text{Re} \left[ p_w(x) \right].
\end{equation}
Here $p_{w}(x) = \xfrac{\langle x\vert p \vert\psi\rangle}{\langle x \vert\psi\rangle}$ is the weak value of the momentum operator \cite{Aharonov1988,Kofman2012,Dressel2014,Tamir2013,Aharonov2014}. If $x$ is a coordinate transverse to the photon's propagation, the dynamics in $x$ is approximately equivalent to a nonrelativistic particle with effective mass $m=\rfrac{h}{\lambda c}$, where $c$ is the speed of light and $\lambda$ is the wavelength. Then we can reconstruct the photon's trajectory using its position and velocity $v(x) =   p(x)/m$,  just as for a classical particle~\cite{KBR2011,MRF2016,YX2017}.

In our experiment we create an effective transverse wave function for the photon that is a superposition of two paths: $\psi^{\phi}(x)= [f_{u}(x)+ e^{i\phi} f_{d}(x)]/\sqrt{2}$. Here $ \phi $ is the relative phase between the paths, while $f_{u(d)}(x) =(w \sqrt{2 \pi})^{-1/2} e^{-(x \mp d/2)^2/ w^2}$
is the Gaussian wave packet, with a waist of $w$, of the upper (lower) path. An ideal WWM creates a two-component entangled state, correlating the path of the particle with orthogonal basis states of a probe. Measuring this probe in a different complementary basis realises a {\em quantum eraser}~\cite{ScuDru82}. In Bohmian mechanics, the details of the momentum change depend on the basis in which the WWM is measured, because of the theory's nonlocality, illustrated in Refs.~\cite{MRF2016,YX2017}. The minimum momentum disturbance occurs when the two outcomes of the quantum eraser measurement project the system state into $\psi^0$ and $\psi^\pi$ respectively~\cite{Wiseman1998}. That is, the particle is still in a superposition of its two paths, just with different relative phases which, when summed, wash out the interference pattern.

We create the above superpositions, and consider an ensemble of Bohmian trajectory starting at $N$
transverse positions  $x_{i}(z_{1})$, where $z_{1}$ represents the initial plane. By reconstructing each trajectory forwards, as described above, to plane $z_{j}$, we can obtain an ensemble of new transverse photon positions $x_{k}^{z_{j}}$ and transverse photon momenta $ p^{\phi}(x_{k}^{z_{j}})$. The change in the Bohmian momentum, for this single trajectory, as a consequence of inducing a phase $\phi \neq 0$ is then calculated as $p_i= p^\phi(x_{k}^{z_{j}}) - p^0(x_{k}^{z_{j}}) $.
By summing over all initial positions $i$, with the appropriate weights, the momentum disturbance distribution can be obtained~\cite{Wiseman1998}:
\begin{align}\label{pro}
 P^\phi_{z_{j}}(p) =  \dfrac{1}{M} \sum_{i=2}^{N-1} & \wp( x_{i}(z_{1}))  %
 \delta(p-p_{i}),
\end{align}
where $ M $ is a factor which ensures $\int_{\infty}^{\infty}  P^\phi_{z_{j}}(p)dp=1$, and
$\wp(x_{i}(z_{1})) = \vert \psi^{\phi}(x_{i}(z_{1})) \vert^{2} (x_{i+1}^{z_{1}} - x_{i-1}^{z_{1}})/2$. In order to obtain a smooth distribution with a finite ensemble size $M$, we approximate
$\delta(p-p_{i})$ by a Gaussian distribution
of standard deviation $\sigma = 0.1 \hbar/d$.
The Bohmian momentum disturbance distribution $ P_{z_j}^{B}(p) $ up to plane $ z_{j} $ due to the WWM can then be calculated as

\begin{equation}\label{distribution}
P_{z_j}^{B}(p)=\eta^{0}P^0_{z_{j}}(p)+\eta^{\pi}P^\pi_{z_{j}}(p),
\end{equation}
where $ \eta^{0}$ $(\eta^{\pi})$ represents the weight of the case $\phi = 0$ $(\phi = \pi)$, with $ \eta^{0}+\eta^{\pi}=1 $. A WWM with perfect path distinguishability corresponds to $\eta^{0}=\eta^{\pi}=1/2$. The {\em total} momentum disturbance distribution, $P_{T}^{B}(p)$, is when $z_j$ is in the far field.

We can quantify the momentum disturbance by
\begin{equation}\label{transfer}
\langle\vert p\vert\rangle_{z_j}^{B} = \int P_{z_j}^{B}(p) |p| dp.
\end{equation}
It was shown theoretically~\cite{Wiseman1998} that, for WWMs
achieving only partial distinguishability, resulting in a nonzero fringe visibility $V$, the  total mean absolute momentum disturbance is bounded below:
\begin{equation} \label{uncertainty}
\langle\vert p\vert\rangle^{B}_{T}\geq \dfrac{2\hbar}{\pi d}(1-V).
\end{equation}
This clearly relates the loss of interference in a WWM to the particle's momentum change. Moreover, the WWM which achieves this bound corresponds to that in \erf{distribution}, with
$\eta^0 = (1+V)/2$.

\begin{figure}[!htbp]
\begin{center}
\includegraphics[width=1\columnwidth]{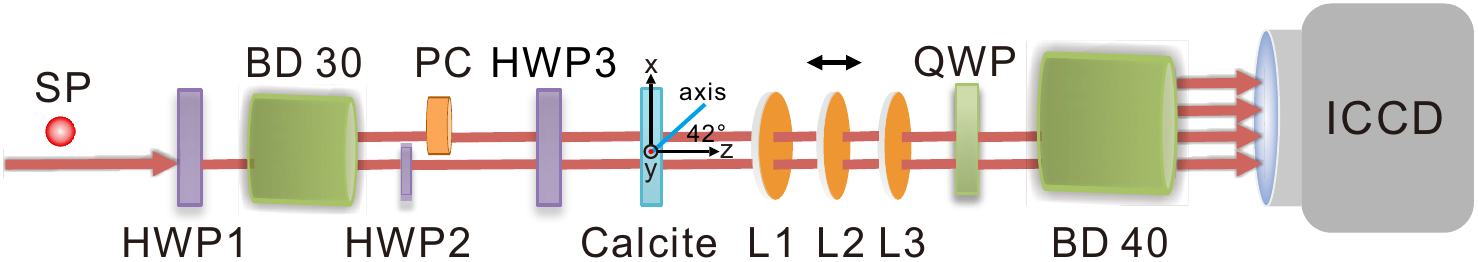}
\caption{Experimental setup.  Heralded signal photons (SPs) are separated into two paths by a beam displacer (BD 30). A half-wave plate HWP1 is used to change the relative phase between these two paths, while HWP2 and HWP3 are used to make the polarization of both paths the same. A birefringent crystal (PC) is inserted into one of the paths to compensate the difference in the optical length. The photon goes through a thin calcite crystal to perform weak measurement. The optic axis of the calcite crystal is in the {\it x-z} plane oriented at $42^{\circ}$ to the $z$ axis. A quarter wave plate (QWP) and a beam displacer (BD 40) are used to detect the polarization of the photon. A combination of three lenses, L1 (plano-convex), L2 (aspherical, moveable) and L3 (plano-convex cylindrical), is used to image different planes on the ICCD camera.}
\label{setup}
\end{center}
\end{figure}

Figure \ref{setup} shows our experimental setup. The generation of heralded signal photons is described in the SI. The signal photon is separated by a beam displacer into its horizontally and vertically polarized components, separated by about 3 mm. By rotating the polarisation of one of these beams, and compensating the difference in their optical paths, they become distinguishable only by their transverse location, describable by a wavefunction $[f_u(x) + e^{i\phi} f_d(x)]/\sqrt{2} $.
 In our experiment, we simulate the quantum eraser WWM by inducing the relative phases $ \phi=0 $ or $ \phi=\pi $,
by rotating HWP1 to $22.5 ^{\circ} $ or $67.5 ^{\circ} $,
respectively.

\begin{figure}[!htb]
\begin{center}
\includegraphics[width=1\columnwidth]{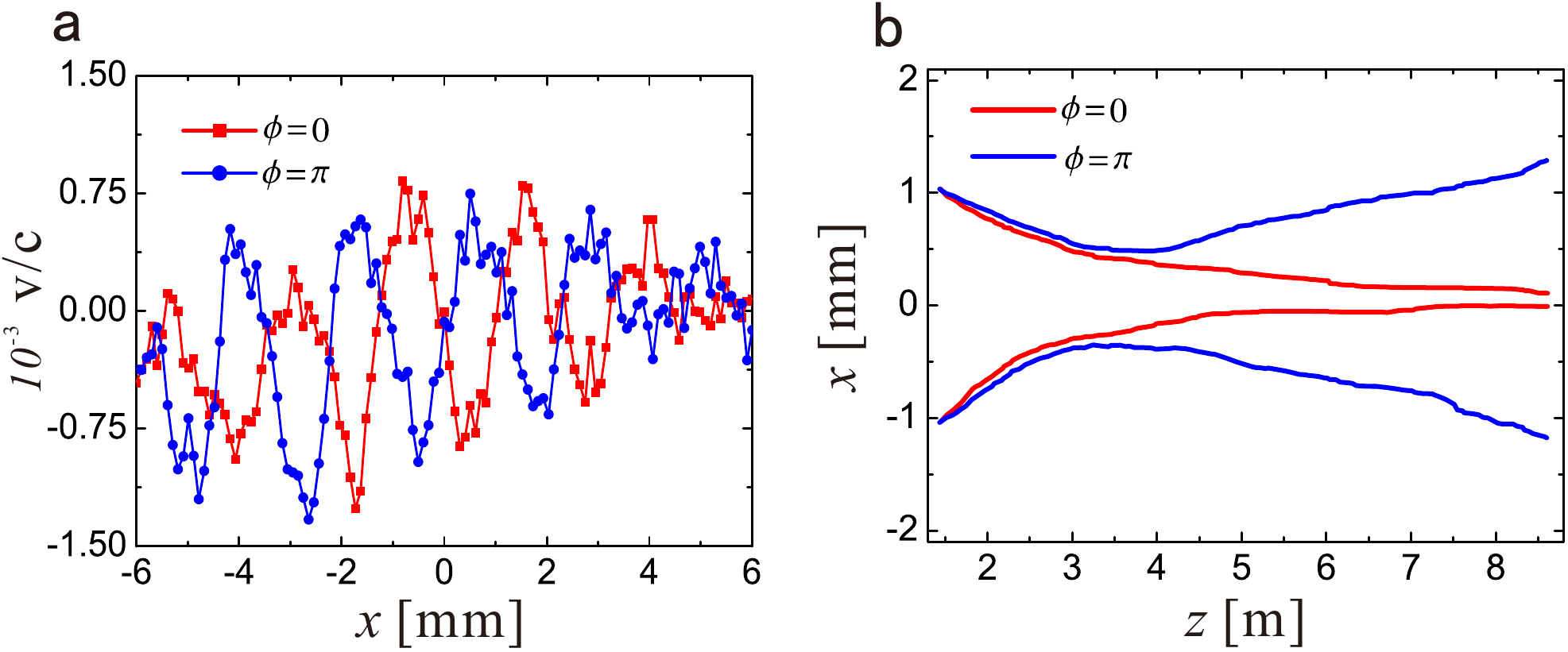}
\caption{Experimental velocities and trajectories. \textbf{a}. The weak value of the transverse velocities or momenta ($v/c = \lambda p/h$) at $z_{117} =8.612$  m . The red squares and blue dots represent experimental data with the relative phase $\phi$ being $ 0 $ and $\pi$, respectively. \textbf{b}. Trajectories beginning at the same initial condition, $x=\pm 1.02$ mm, for $\phi= 0 $ (red) and $\phi= \pi $ (blue). The trajectories are reconstructed from 117 imaging planes. If a trajectory locates on a point that is not at the center of a pixel, then a cubic spline interpolation between neighbouring momentum values is used.}
\label{traj}
\end{center}
\end{figure}

The signal photon is then sent to the transverse momentum (or velocity) measurement setup, which consists of a 0.7-mm-thick piece of calcite with its optic axis in the {\it x-z} plane oriented at $42^{\circ}$ to the $z$ axis followed by a quarter wave-plate and a beam displacer. The photon's position  $x_{k}^{z_{j}}$  at the $ z_{j} $ plane is recorded by the pixels of the ICCD camera (Andor iStar 334), which is triggered by the electronic signal from the detection of the idler photon. The weak value of the transverse momentum is obtained from many runs by
\beq
\langle p(x_{k}^{z_{j}})\rangle_{w}=\dfrac{h}{\lambda\zeta} \arcsin\left(\dfrac{N^{j}_{R_{k}}-N^{j}_{L_{k}}}{N^{j}_{R_{k}}+N^{j}_{L_{k}}}\right),
\eeq
where 
$1/\zeta =1/336 $ is the measured dimensionless coupling strength~\cite{YX2017} and $N^{j}_{R_{k} (L_{k})}$ is the photon count corresponding to the right-hand (left-hand) circular polarization. From this we obtain the transverse Bohmian velocity as $v(x_{k}^{z_{j}}) = \langle p(x_{k}^{z_{j}})\rangle_{w}/m$. A system of three lenses (L1, L2 and L3) with the middle lens translatable in the $z$ direction, allows us to vary $z_j$ 
from the near field $z_1=1.445$ m) to the far field ($z_{117}=8.612$ m). With the initial transverse position and the velocity at different imaging plane, we reconstruct the photon's average trajectories according to $ x_{k}^{z_{j+1}}=x_{k}^{z_j}+{(z_{j+1}-z_j)} v(x_k^{z_{j}}) /\sqrt{c^{2}-v^{2}(x_k^{z_j})} $, as in Ref.~\cite{YX2017}, which is practically identical to $x_{k}^{z_j}+{(z_{j+1}-z_j)} p(x_{k}^{z_j})/mc$.

In the experiment, we consider 198 initial positions  $ x_{i}(z_{1})$, with 99 for each slit. The initial positions are chosen to equally sample the Gaussian distribution of $|\psi(x)|^2$ across each slit (see Methods for details).
The weak value of the transverse momentum at $z_{117} =8.612$ m is shown in Fig.~\ref{traj}\textbf{a}, as a function of $x$ and for both phases $\phi$.
The relative phase difference of $\pi$ yields the complementary pattern in momenta.
Fig. \ref{traj}\textbf{b} shows the reconstructed trajectories beginning at the same place (two places are chosen: $x=\pm 1.02$ mm) for the two phases. For $\phi=0$, the trajectories converge to form the zero-order fringe, while for $\phi=\pi$, they diverge towards the two first-order fringes.

\begin{figure}[!htb]
\begin{center}
\includegraphics[width=0.9\columnwidth]{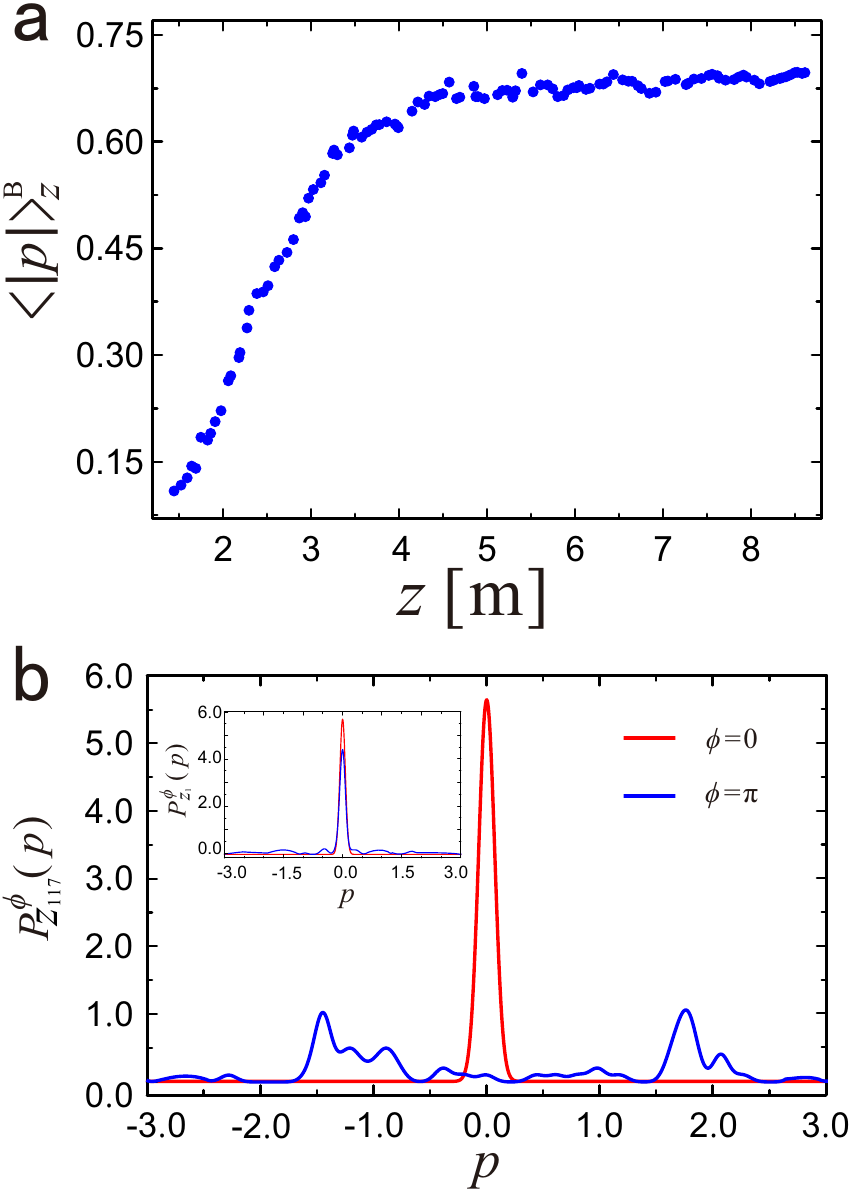}
\caption{Bohmian momentum disturbance, in units of $\hbar/d$, for the photons subject to a minimally disturbing WWM with $\eta^{0}=\eta^{\pi}=1/2$. \textbf{a}. The mean absolute  momentum disturbance $ \langle\vert p\vert\rangle^{B}_{z}$ as a function of the propagation distance $z$.  The nonclassicality of the momentum change is evident from the fact that it accrues gradually during propagation from near to far field. \textbf{b}. Complete distribution of the Bohmian momentum disturbance at the last plane and the first plane (inset). Red is for $\phi=0$ and blue for $\phi=\pi$.
}
\label{probability}
\end{center}
\end{figure}

Following all the trajectories, we can obtain the Bohmian momentum disturbance distribution $ P^{\phi}_{z_j}(p) $ and the mean absolute momentum disturbance $\langle\vert p\vert\rangle^{B}_{z_j}$ at different planes $z_j$  using Eq. (\ref{pro}) and Eq. (\ref{transfer}), respectively. The latter as a function of $z$ is shown in Fig. \ref{probability}\textbf{a} with $\eta^{0}=\eta^{\pi}=1/2$. This shows that the momentum disturbance is not a momentum kick that occurs at the point of the WWM~\cite{Wiseman1998}. Rather, it gradually accumulates during the propagation of the photons. This delayed Bohmian momentum disturbance is characteristic of a nonclassical momentum disturbance as defined in Ref.~\cite{Wiseman1995}.

To further demonstrate the difference, Fig.~\ref{probability}\textbf{b} compares the total Bohmian momentum disturbance distribution $P^{\phi}_{z_j}(p)$, at the last plane ($z_{117}=8.612$ m), with that at the first plane ($z_{1}=1.445$ m) (inset). The two peaks in $P^{\pi}_{z_{117}}(p)$ at $p\approx\pm\rfrac{3\hbar}{2 d}$ are the dominant contribution to the mean absolute Bohmian momentum disturbance. They are almost absent from $P^{\pi}_{z_{1}}(p)$, since in the near field there has not been sufficient time for the wavefunction to develop the different phase gradients for the different $\phi$ that guide the Bohmian photons. The lack of any immediate disturbance to the Bohmian momentum from our WWM reflects the fact that the moments of the far-field momentum distribution are unchanged by this type of WWM~\cite{Wiseman1997,Wiseman1998,Wiseman2003}.



\begin{figure}[!htb]
\begin{center}
\includegraphics[width=0.80\columnwidth]{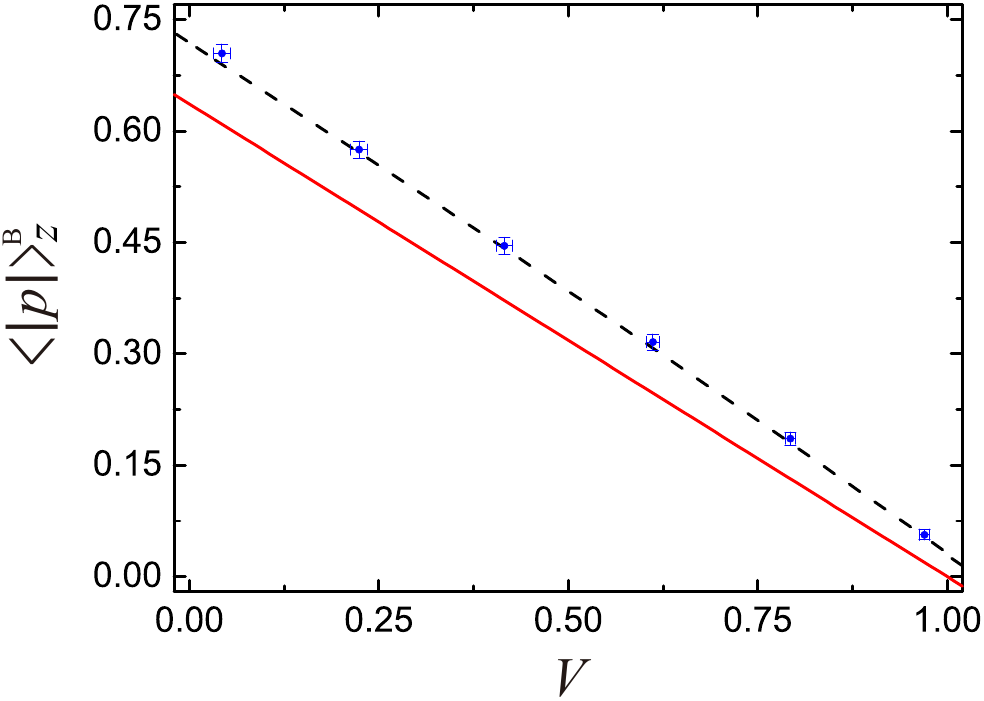}
\caption{The relationship between the total mean absolute momentum disturbance $\langle\vert p\vert\rangle^{B}_{z}$ (in units of $ \hbar/d $) and fringe visibility $V$. The blue dots are experimental data for various partial WWMs kinds of measurements in the plane $z_{117}=8.612$ m  . The red solid line is the theoretical prediction (\ref{uncertainty}) under ideal conditions. The black dashed line represents the theoretical prediction, calculated with the same experimental conditions. Error bars are estimated from the counting statistics.}
\label{visibility}
\end{center}
\end{figure}

We also look at the tradeoff between the  mean absolute momentum disturbance $\langle\vert p\vert\rangle^{B}_{T}$ and the visibility $V$ by changing the relative weight of the cases of $\phi=0$ and $\phi=\pi$. This corresponds to WWMs with partial information, allowing a nonzero visibility $V$ to remain.  The results are shown in Fig. \ref{visibility}.
The increase in the momentum disturbance as the visibility is reduced is clearly observed, 
and always exceeds the theoretical bound of inequality~(\ref{uncertainty}). The methods to estimate the fringe visibility $ V $  can be found in Methods.

For the WWM we implement, this bound should be achievable. However, this bound is calculated considering infinitely many initial positions at the the plane $z=0$, whereas in our experiment we have 198 initial positions at the plane $z_{1}=1.445$ m.
Thus to compare with our experiment we also calculate  $V$ and, in the framework of Bohmian mechanics, $\langle\vert p\vert\rangle^{B}_{z_{117}}$, but using the Gaussian approximation to the delta-function to get $P^B_{z_{117}}(p)$ and using the same experimental conditions (117 image planes from $z_{1}=1.445$ m to $z_{117}=8.612$ m with 198 initial positions). The result is the black dashed line 
and now agrees well with the experimental results.

In this work, we used the Bohmian probability distribution~\cite{Wiseman1998} to experimentally quantify the momentum disturbance arising from a WWM which destroys (or partially destroys) two-slit interference. In particular, we measured the mean of the absolute value of the Bohmian momentum disturbance $\langle\vert p\vert\rangle^{B}_{z_j}$. In the far field, the results we obtained are consistent with the theoretical inequality $ \langle\vert p\vert\rangle^{B}_{T}\geq (1-V)\rfrac{2\hbar}{\pi d}$, where $V$ is the remaining visibility when the WWM is only partial. We also show that $\langle\vert p\vert\rangle^{B}_{z_j}$ in the near field is small, and acquires its far-field value gradually as the photon travels longitudinally. This is characteristic~\cite{Wiseman1998} of a momentum disturbance which is nonclassical~\cite{Wiseman1995,AhaPenPet69}.

Finally, we note that there are other methods to characterise the momentum transfer~\cite{Wiseman1997,Wiseman2003} --- the latter having been realised experimentally~\cite{Wiseman2007} ---
which also reflect the
difference between classical and nonclassical momentum disturbance.
While the momentum disturbance distributions in these methods have the advantage of being independent of the basis used for reading out the WWM device, they are not true probability distributions: they take negative values for nonclassical cases.
By contrast, in our experiment we measured a family of true probability distributions, which quantitatively capture the relationship between momentum disturbance and fringe visibility, and which also enabled us to show the nonclassicality of that disturbance quantitatively for the first time. Thus, treating the momentum as an element of reality in Bohmian mechanics arguably provides the most useful method to understand the change of photon's momentum in a WWM.
 Moreover, it gives an intuitive picture of part of the ``uncontrollable change in the momentum''~\cite{Heisenberg1927} which enforces complementarity: although Bohmian dynamics is fully deterministic, the momentum transfer experienced by a particle depends on its initial position within the wavefunction, and that cannot be controlled by the experimenter.


\section{Methods}
{\bf The selection of initial transverse positions $x_{i}(z_{1})$:}
The probability distributions of the signal photon in the up and down paths at $ z_{1} $ plane are represented as $P_{u}(x)=\vert f_{u}(x)\vert^{2}$ and $P_{d}(x)=\vert f_{d}(x)\vert^{2}$, respectively.  198 trajectories are reconstructed in the experiment with 99  for each beam.
The initial transverse positions $x_{i}(z_{1})$ of each beam are chosen to satisfy $A_{i}=\int_{-\infty}^{x_{i}(z_{1})} P_{u(d)}(x) dx=\xfrac{i}{100}$ with $ i\in\lbrace 1, 2...99 \rbrace $.

{\bf Estimation of the interference visibility $V$:} The intensities detected in the ICCD camera are denoted as $N_{k}^{0}(z_{j})$ and $N_{k}^{\pi}(z_{j})$ at the position $x_k$ at $z_{j}$ plane for the relative phase being $0$ and $\pi$, respectively. The probabilities of these two cases are denoted as $\eta^{0}$ and $\eta^{\pi}$, respectively. The total intensity at $ z_{j} $ plane can be calculated as $N_{k}(z_{j}) = \eta^{0}N_{k}^{0}(z_{j})+\eta^{\pi} N_{k}^{\pi}(z_{j}) $ with $\eta^{0}+\eta^{\pi}=1$. The minimum intensities of the first-order interference are denoted as $N_{1}$ and $N_{2}$, respectively. The maximum intensity of the zero-order interference is denoted as $N_{3}$. The visibility $ V $ is then calculated as $ V=\dfrac{N_{3}-(N_{1} + N_{2})/2}{N_{3}+(N_{1} + N_{2})/2}. $

For the comparison between experiment and theory (the black dashed line in Fig. 4), the visibility is further theoretically estimated with the same experimental conditions in the framework of quantum theory. The intensity distribution is calculated as $I=\eta^{0}\vert\psi^{0}(x_{k}^{z_{j}})\vert^{2}+\eta^{\pi}\vert\psi^{\pi}(x_{k}^{z_{j}})\vert^{2}$. The visibility $V$ can then be obtained as $V=(I_{\rm max}-I_{\rm min})/(I_{\rm max}+I_{\rm min})$, where $I_{\rm max}$ and $I_{\rm min}$ represent the zero-order maximum intensity and first-order minimum intensity, respectively.


\section{Supplementary Information}

\subsection{Momentum in Bohmian Mechanics and Related Theories}

In this paper, we are taking Bohmian mechanics to be defined by the three assumptions on page 171 of Ref.~\cite{Bhom1}, the second of which is
\begin{quote}
(2) That the particle momentum is restricted to ${\bf p} = \nabla S({\bf x})$.
\end{quote}
Here, $S/\hbar$ is the phase of the wavefunction.
Thus the momentum does not have autonomous dynamics but rather is determined by the position ${\bf x}$ (and the wavefunction). For this reason, some reformulations of Bohmian mechanics take only the position to be fundamental, and downgrade the momentum to an emergent element equal (in the nonrelativistic case) to mass times the rate of change of the position~\cite{DGZ13,Val97}.  Alternatively, Bohm's momentum can be considered the ``local expectation value'' of momentum~\cite{Holland93}, just like the local value of spin addressed experimentally in Ref.~\cite{MRF2016}.

\subsection{Single Photon Generation}
\begin{figure}[!htbp]
\begin{center}
\includegraphics[width=1\columnwidth]{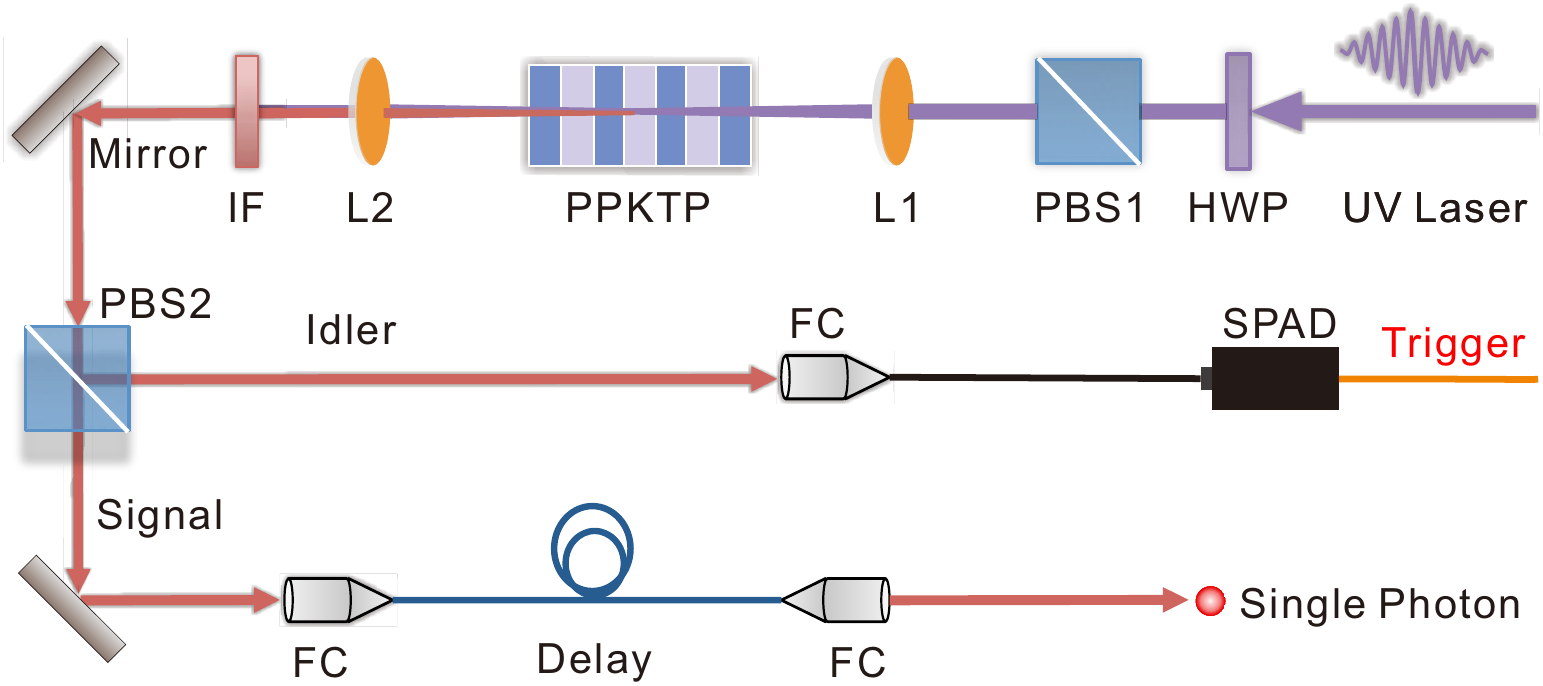}
\caption{Experimental setup for single photon generation.  The intensity and polarization of the ultraviolet laser are set by a half-wave plate (HWP) and a polarization beam splitter (PBS1). The laser is focused on a periodically poled KTiPO4 (PPKTP) crystal via a lens (L1). The down-converted photons are collimated with another lens (L2) and filtered by an interference filter (IF).
The signal and idler photons are separated by PBS2 with the idler photon coupled by a fiber coupler (FC) and detected by a single-photon avalanche detector (SPAD) with the electronic signal using as a trigger. The signal photon is delayed by a $85$ m long single-mode fiber to meet the detecting time in the ICCD camera (shown in Fig. 1 in the main text).}
\end{center}
\end{figure}

A $404$ nm laser (Toptica Bluemode) is used to pump a $20$ mm long periodically poled KTiPO4 (PPKTP) crystal and down-converted photons are filtered by an interference filter centered at $808$ nm  with a bandwidth of $3 $ nm (LL01-808, Semrock). The spectral brightness is obtained to be $ 50,000 $ pairs/(s mW) when the PPKTP crystal is operated at the degenerate temperature of $ 31.00 \pm 0.01^{\circ}$C. After separated by a polarization beam splitter (PBS2), the idler photon is directly detected by a single-photon avalanche detector with the electric signal using as a trigger. The signal photon is first delayed by a 85 m long single-mode fiber to make sure the detection is made in time and then sent to a polarization dependent two-slit setup shown in Fig. 1 in the main text.


\section{Acknowledgments}
This work was supported by the National Key Research and Development Program of China (Grants No. 2016YFA0302700 and 2017YFA0304100), the National Natural Science Foundation of China (Grants No.\ 61725504, 61327901 and 11774335), Anhui Initiative in Quantum Information Technologies (Grants No. AHY060300 and AHY020100), the Key Research Program of Frontier Sciences, CAS (Grants No. QYZDY-SSW-SLH003), the Fundamental Research Funds for the Central Universities (Grants No.\ WK2470000020 and WK2470000026).

%



\end{document}